\documentclass[aps,prd,preprint,amsmath,amssymb,floatfix]{revtex4}

\usepackage{graphicx}

\begin{document}

\title{Probe of unparticles at the LHC in exclusive two lepton and two photon production via photon-photon fusion}

\author{\.{I}. \c{S}ahin}
\email[]{inancsahin@karaelmas.edu.tr}
 \affiliation{Department of Physics, Zonguldak Karaelmas University, 67100 Zonguldak, Turkey}

\author{S. C. \.{I}nan}
\email[]{sceminan@cumhuriyet.edu.tr} \affiliation{Department of
Physics, Faculty of Sciences, Ankara University, 06100 Tandogan,
Ankara, Turkey} \affiliation{Department of Physics, Cumhuriyet
University, 58140 Sivas, Turkey}

\begin{abstract}
The exclusive production $pp\to pXp$ is known to be one of the most
clean channels at the LHC. We investigate the potential of processes
$pp\to p\ell^-\ell^+p$ and $pp\to p\gamma\gamma p$ to probe scalar
and tensor unparticles by considering three different forward
detector acceptances; $0.0015<\xi<0.15$, $0.0015<\xi<0.5$ and
$0.1<\xi<0.5$. We obtain 95\% confidence level sensitivity limits on
the unparticle couplings for various integrated luminosities.
\end{abstract}

\maketitle

\section{Introduction}
The Large Hadron Collider (LHC) generates high energetic
proton-proton collisions with a luminosity of ${\cal L}=10^{34}
cm^{-2}s^{-1}$. It provides high statistics data at high energies.
On the other hand hadronic interactions generally involve serious
backgrounds which have to be managed. Recently a new phenomenon
called exclusive production was observed in the measurements of CDF
collaboration \cite{cdf1,cdf2,cdf3,cdf4,cdf5,cdf6,cdf7} and its
physics potential has being studied at the LHC
\cite{lhc1,lhc2,lhc3,lhc4,lhc5,lhc6,albrow,lhc7}. Complementary to
proton-proton interactions, studies of exclusive production of
leptons and heavy particles might be possible and opens new field of
studying very high energy photon-photon and photon-proton
interactions.

The exclusive production $pp\to pXp$, provides a clean environment
due to absence of the proton remnants. ATLAS and CMS collaborations
have a program of forward physics with extra detectors located in a
region nearly 100m-400m from the interaction point. These forward
detector equipment allows us to detect intact scattered protons
after the collision. Therefore the processes which spoil the proton
structure, can be easily discerned from the exclusive
photo-production processes. By use of forward detector equipment we
can eliminate many serious backgrounds. This is one of the
advantages of the exclusive photo-production processes.  Moreover
photon-induced reactions are electromagnetic in nature and due to
absence of the proton remnants it is free from almost all
backgrounds. One possible background is the proton dissociation into
baryon excitations. But this background can be eliminated
effectively by imposing a cut on the transverse momentum of the
photon or lepton pair \cite{lhc3}. It was argued in \cite{lhc3} that
photon-induced lepton pair production is one of the most clean
channels at the LHC when the acceptance cuts in place.

In this work we investigate the potential of exclusive $pp\to
p\ell^-\ell^+p$ and $pp\to p\gamma\gamma p$ reactions at the LHC to
probe unparticles. Unparticles are non-integral number $d_U$ of
particles. They are manifestations of a possible scale invariant
sector of the new physics that may interact weakly with the standard
model (SM) fields \cite{georgi1,georgi2,cheung1}. At low energies
several effective interaction terms between unparticles and SM
particles can be considered. In our calculations we consider the
following effective interaction operators between SM fields and
unparticles that satisfy the SM gauge symmetry \cite{cheung2}:

\begin{eqnarray}
\label{scalarop}
 \frac{\lambda_S}{\Lambda_{U}^{d_{U}-1}}\bar{f}f
{\cal O_U},\,\,\,\,\,
\frac{\lambda_{PS}}{\Lambda_{U}^{d_{U}-1}}\bar{f}i\gamma^5 f {\cal
O_U},\,\,\,\,\,\frac{\lambda_V}{\Lambda_{U}^{d_{U}}}\bar{f}\gamma^{\mu}f(\partial_{\mu}{\cal
O_U}) ,\,\,\,\,\,\frac{\kappa}{\Lambda_{U}^{d_{U}}}
G_{\mu\nu}G^{\mu\nu}{\cal O_U}
\end{eqnarray}

\begin{eqnarray}
\label{tensorop}
-\frac{1}{4}\frac{\lambda_2}{\Lambda_{U}^{d_{U}}}\bar{\psi}i(\gamma_{\mu}D_{\nu}
+\gamma_{\nu}D_{\mu})\psi {\cal O_{U}^{\mu\nu}},\,\,\,\,\,\,
\frac{\lambda_2^\prime}{{\Lambda_{U}^{d_{U}}}}G_{\mu\alpha}
G_{\nu}^{\alpha}{\cal O_{U}^{\mu\nu}}
\end{eqnarray}
where
$D_\mu=\partial_\mu+ig\frac{\tau^{a}}{2}W_{\mu}^{a}+ig'\frac{Y}{2}B_{\mu}$
is the covariant derivative, $G^{\alpha\beta}$ denotes the gauge
field strength. $f$ stands for a SM fermion and $\psi$ is the SM
fermion doublet or singlet. ${\cal O_U}$ and ${\cal O_{U}^{\mu\nu}}$
represent the scalar and tensor unparticle fields. Feynman rules for
these operators were given in \cite{cheung2}.

Two-point functions for unparticles can be obtained by imposing
scale invariance (or conformal invariance)
\cite{georgi1,georgi2,cheung1,grinstein}. Requiring scale
invariance, the Feynman propagators for the scalar and tensor
unparticles are given respectively by

\begin{eqnarray}
\label{scalarprop}
\Delta(P^{2})=i\frac{A_{d_{U}}}{2sin(d_{U}\pi)}(-P^{2})^{d_{U}-2}
\end{eqnarray}
\begin{eqnarray}
\label{tensorprop}
\Delta(P^{2})_{\mu\nu,\rho\sigma}=i\frac{A_{d_{U}}}{2sin(d_{U}\pi)}(-P^{2})^{d_{U}-2}
T_{\mu\nu,\rho\sigma}(P)
\end{eqnarray}
where,
\begin{eqnarray}
A_{d_{U}}=\frac{16\pi^{\frac{5}{2}}}{(2\pi)^{2d_{U}}}\frac{\Gamma(d_{U}+\frac{1}{2})}
{\Gamma(d_{U}-1)\Gamma(2d_{U})}
\end{eqnarray}
\begin{eqnarray}
T_{\mu\nu,\rho\sigma}(P)=\frac{1}{2}\left[\pi_{\mu\rho}(P)\pi_{\nu\sigma}(P)+
\pi_{\mu\sigma}(P)\pi_{\nu\rho}(P)-\frac{2}{3}\pi_{\mu\nu}(P)\pi_{\rho\sigma}(P)\right]
\end{eqnarray}
\begin{eqnarray}
\pi_{\mu\nu}(P)=-g_{\mu\nu}+\frac{P_{\mu}P_{\nu}}{P^{2}}
\end{eqnarray}

Conformal invariance can also be used to fix unparticle two-point
functions. Conformal invariance leads to the same propagator for the
scalar unparticles. But the tensor unparticle propagator is modified
to a different form \cite{grinstein}. In
Refs.\cite{grinstein,nakayama} theoretical bounds on the scale
dimension were obtained from unitarity constraints. The scaling
dimension for the scalar unparticle is constrained as $d_U\geq1$.
This constraint is valid in both conformal and scale invariance.
Scale invariance restricts the scaling dimension of tensor
unparticle operator to $d_U\geq3$. On the other hand, conformal
invariant imposes a constraint of $d_U\geq4$. We do not consider
conformal invariance in the case of tensor unparticles since the
lower bound of the scale dimension is large and therefore unparticle
contribution is very suppressed. But we will present some results
for the scale invariant tensor unparticles with the scale dimension
$d_U=3.001$ and $d_U=3.01$.

\section{Equivalent Photon Approximation and photon-photon fusion}

The photon-photon fusion can be described by equivalent photon
approximation (EPA) \cite{budnev}. In the exclusive production of an
object X, two photons scattered from protons interact each other
through the process $pp\to p\gamma\gamma p\to pXp$. In the framework
of EPA, emitted photons have a low virtuality  and scattered with
small angles from the beam pipe. Therefore they are almost real and
the cross section for the complete process $pp\to p\gamma\gamma p\to
pXp$ can be obtained by integrating the cross section for the
subprocess $\gamma \gamma \to X$ over the effective photon
luminosity $\frac{d L^{\gamma \gamma}}{dW}$

\begin{eqnarray}
\label{completeprocess}
 d\sigma=\int{\frac{dL^{\gamma\gamma}}{dW}}
\,d\hat {{\sigma}}_{\gamma\gamma \to X}(W)\,dW
\end{eqnarray}
where W is the invariant mass of the two photon system and the
effective photon luminosity is given by
\begin{eqnarray}
\label{efflum}
\frac{dL^{\gamma\gamma}}{dW}=\int_{Q^{2}_{1,min}}^{Q^{2}_{max}}
{dQ^{2}_{1}}\int_{Q^{2}_{2,min}}^{Q^{2}_{max}}{dQ^{2}_{2}} \int_{y_{
min}}^{y_{max}} {dy \frac{W}{2y} f_{1}(\frac{W^{2}}{4y}, Q^{2}_{1})
f_{2}(y,Q^{2}_{2})}.
\end{eqnarray}
with
\begin{eqnarray}
y_{min}=\mbox{MAX}(W^{2}/(4\xi_{max}E), \xi_{min}E), \;\;\;
y_{max}=\xi_{max}E.
\end{eqnarray}
$Q_{max}^2$ is taken to be $2\,\text{GeV}^2$, $y$ is the energy of
one of the emitted photons from the proton, $\xi_{min}$ and
$\xi_{max}$ are the acceptances of the forward detectors which tag
protons with some momentum fraction loss
$\xi=(|\vec{p}|-|\vec{p}^{\,\,\prime}|)/|\vec{p}|$. $f_1$ and $f_2$
are the equivalent photon spectra. Equivalent photon spectrum of
virtuality $Q^2$ and energy $E_\gamma$ is given by

\begin{eqnarray}
f=\frac{dN}{dE_{\gamma}dQ^{2}}=\frac{\alpha}{\pi}\frac{1}{E_{\gamma}Q^{2}}
[(1-\frac{E_{\gamma}}{E})
(1-\frac{Q^{2}_{min}}{Q^{2}})F_{E}+\frac{E^{2}_{\gamma}}{2E^{2}}F_{M}]
\end{eqnarray}
where
\begin{eqnarray}
Q^{2}_{min}=\frac{m^{2}_{p}E^{2}_{\gamma}}{E(E-E_{\gamma})},
\;\;\;\; F_{E}=\frac{4m^{2}_{p}G^{2}_{E}+Q^{2}G^{2}_{M}}
{4m^{2}_{p}+Q^{2}} \\
G^{2}_{E}=\frac{G^{2}_{M}}{\mu^{2}_{p}}=(1+\frac{Q^{2}}{Q^{2}_{0}})^{-4},
\;\;\; F_{M}=G^{2}_{M}, \;\;\; Q^{2}_{0}=0.71 \mbox{GeV}^{2}
\end{eqnarray}
Here E is the energy of the proton beam which is related to the
photon energy by $E_{\gamma}=\xi E$ and $m_{p}$ is the mass of the
proton. The magnetic moment of the proton is taken to be
$\mu^{2}_{p}=7.78$. $F_{E}$ and $F_{M}$ are functions of the
electric and magnetic form factors.

\begin{figure}
\includegraphics{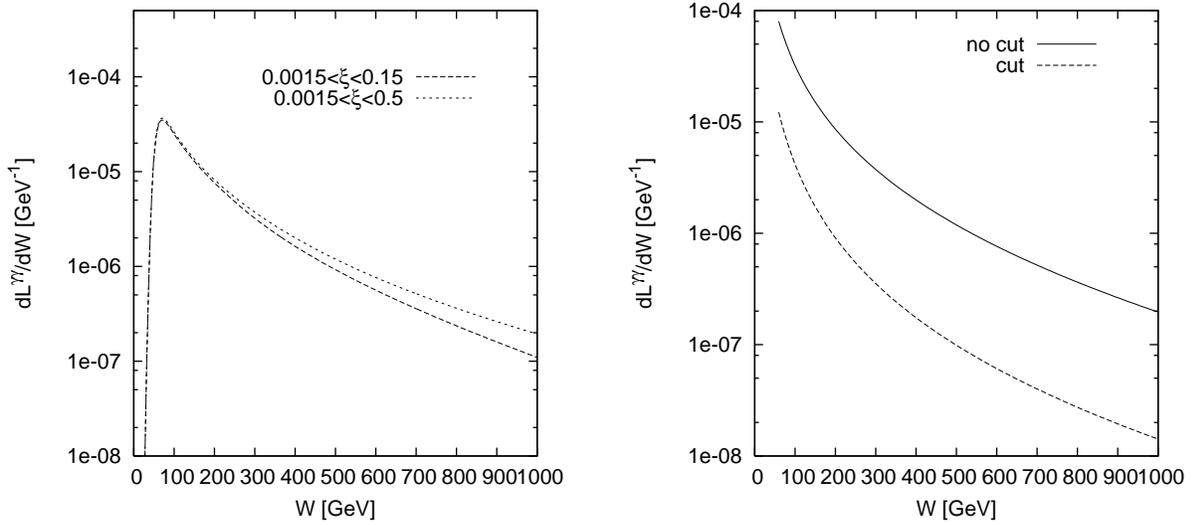}
\caption{Effective $\gamma\gamma$ luminosity as a function of the
invariant mass of the two photon system. Figure on the left shows
effective luminosity for forward detector acceptances
$0.0015<\xi<0.15$ and $0.0015<\xi<0.5$. Figure on the right
represents the cases with and without a cut on transverse momentum
of the photon pair $|\vec{q}_{1t}+\vec{q}_{2t}|<30$MeV. In the right
panel, we do not consider any acceptance i.e., $\xi$ is taken to be
in the interval $0<\xi<1-m_p/E$ where $m_p$ is the mass and $E$ is
the energy of the incoming proton. \label{fig1}}
\end{figure}

The object X is detected by the central detectors while the intact
scattered protons are detected by the forward detectors. ATLAS and
CMS have central detectors with a pseudorapidity coverage
$|\eta|<2.5$. ATLAS Forward Physics (AFP) Collaboration proposed an
acceptance of $0.0015<\xi<0.15$ \cite{albrow}. This acceptance
allows to detect an object of mass in the interval $100\,
GeV<M<800\, GeV$ with a good accuracy. There are also other
scenarios with different acceptances of the forward detectors.
CMS-TOTEM forward detector scenario spans $0.0015<\xi<0.5$ and
$0.1<\xi<0.5$ \cite{avati,lhc6}. In Fig.\ref{fig1} in the left
panel, we plot effective $\gamma\gamma$ luminosity as a function of
invariant mass of the two photon system for various forward detector
acceptances.

In Ref.\cite{lhc3}, exclusive lepton-pair production via photon
photon fusion was proposed as a luminosity monitor for the LHC. It
was discussed in detail in \cite{lhc3} that main possible background
is the proton dissociation into baryon excitations; $pp \to
X+\ell^{+}\ell^{-}+Y $ where X and Y are baryon excitations such as
$N^{*}$, $\Delta$ isobars. It was shown in \cite{lhc3} that this
background can be eliminated effectively by imposing a cut on the
transverse momentum of the photon pair
$|\vec{q}_{1t}+\vec{q}_{2t}|<(10-30)$MeV. In actual experiment this
cut can be placed on either photon pair or lepton pair. Similar
arguments is also true for exclusive two photon production and same
cut should be applied in order to eliminate the contamination from
proton dissociation into baryon excitations. In all the results
presented in this work we impose a cut of
$|\vec{q}_{1t}+\vec{q}_{2t}|<30$MeV on the transverse momentum of
the photon pair. To see the effect of this cut on the effective
$\gamma\gamma$ luminosity, we plot $dL^{\gamma\gamma}/dW$ as a
function of invariant mass of the two photon system with and without
a cut in the right panel of Fig.\ref{fig1}.

\section{cross sections and numerical analysis}
\subsection{Exclusive two lepton production}

\begin{figure}
\includegraphics{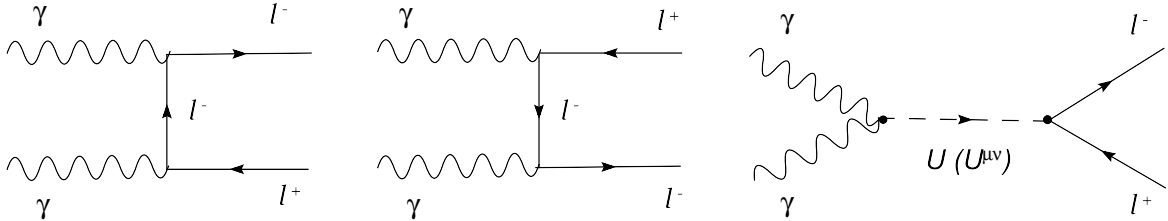}
\caption{Tree-level Feynman diagrams for the subprocess
$\gamma\gamma \to \ell^-\ell^+$. \label{fig2}}
\end{figure}

In the SM, the subprocess $\gamma\gamma \to \ell^-\ell^+$ is
described by t and u-channel tree-level diagrams. New physics
contribution comes from s-channel unparticle exchange
(Fig.\ref{fig2}). The polarization summed amplitude square is given
by the following formula

\begin{eqnarray}
\label{ggll}
|M|^2=&&8g_e^4tu(\frac{1}{t^2}+\frac{1}{u^2})+\frac{4A_{d_{U}}^2s^{(2d_u-4)}}{\sin^2(d_u\pi)}
\frac{\kappa^2}{\Lambda_U^{(4d_U-2)}}(\lambda_{PS}^2 + \lambda_S^2)s^3 \nonumber \\
&&+\frac{A_{d_{U}}^2s^{(2d_u-4)}}{2\sin^2(d_u\pi)}
\left(\frac{\lambda_2^2 \lambda_2^{\prime\,2}}{\Lambda_U^{4d_U}}\right)ut(t^2 + u^2) \nonumber \\
&&-4g_e^2A_{d_{U}}s^{(d_u-2)}\left(\frac{\lambda_2\lambda_2^{\prime}}{\Lambda_U^{2d_U}}\right)\cot(d_u\pi)
(t^2+u^2)
\end{eqnarray}
where $g_e=\sqrt{4\pi\alpha}$, s,t and u are the Mandelstam
variables and we omit the mass of leptons. We see from this
amplitude that scalar unparticle contribution does not interfere
with the SM. Therefore scalar unparticle contribution is always
additive. On the other hand, tensor contribution interfere with the
SM. The trigonometric functions $\cos(d_u\pi)$ in the interference
terms originate from the complex phase associated with the s-channel
propagator and may lead to interesting interference effects with the
standard model amplitudes. We also see from (\ref{ggll}) that
contribution of the coupling $\lambda_{PS}$ to the cross section is
equal to the contribution of the coupling $\lambda_{S}$. It is then
impossible to distinguish $\lambda_{PS}$ from $\lambda_{S}$ and
therefore we only consider the coupling $\lambda_{S}$ in our
numerical calculations. The scalar unparticle coupling $\lambda_V$
does not contribute to the process since the unparticle couples to
the on-mass-shell current $\ell^-\ell^+$.

\begin{figure}
\includegraphics{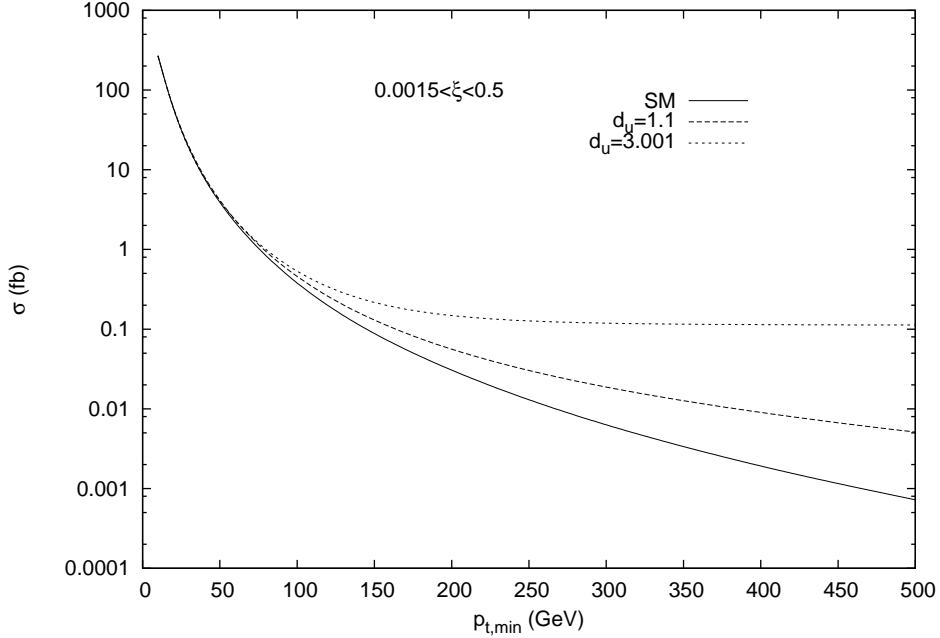}
\caption{Cross section of $pp\to p\ell^-\ell^+p$ as a function of
the transverse momentum cut on the final leptons. Solid line is for
the SM and dotted lines $d_u=1.1$ and $d_u=3.001$ include scalar and
tensor unparticle contributions respectively. Scalar unparticle
couplings are taken to be $\kappa=\lambda_S=1$ and tensor unparticle
couplings are taken to be $\lambda_2^\prime=1$ and $\lambda_2=10^3$.
\label{fig3}}
\end{figure}

We consider three different forward detector acceptances;
$0.0015<\xi<0.15$, $0.0015<\xi<0.5$ and $0.1<\xi<0.5$. In
Fig.\ref{fig3} we plot cross section of $pp\to p\ell^-\ell^+p$ as a
function of the transverse momentum cut on the final leptons. We
observe from the figure that cross sections including unparticle
contributions deviate from the SM as the $p_t$ cut increases.
Unparticle contributions and the SM  are well separated from each
other for large values of the $p_t$ cut. Furthermore, we observe
from (14) that the SM contribution is highly peaked in the forward
and backward directions due to $t,u=0$ poles whereas the unparticle
contribution is rather flat. Therefore both angular distribution or
the $p_t$ cut can be used to improve sensitivity bounds.

During statistical analysis we use two different approach. In the
first approach we impose cuts on the transverse momentum of the
final leptons to suppress the SM cross section. We make the number
of SM event less than 0.5. Then it is very appropriate to set bounds
on the couplings using a Poisson distribution. We set a cut of
$p_t>420$ GeV for $0.0015<\xi<0.15$ and a cut of $p_t>460$ GeV for
$0.0015<\xi<0.5$ on the final leptons to improve the bounds. These
values for the $p_t$ cut make the SM event less than 0.5 for a
luminosity of 200 $fb^{-1}$.  In the case $0.1<\xi<0.5$, invariant
mass of the final leptons is greater than 1400 GeV due to the high
lower bound of $\xi$. The SM cross section is very small and
therefore it does not need to impose a high $p_t$ cut. We consider a
cut of $p_t>30$GeV for $0.1<\xi<0.5$.

In the second approach we have obtained sensitivity bounds using the
simple $\chi^2$ criterion from angular distribution

\begin{eqnarray}
\chi^2=\sum_{i=bins}\left(\frac{\sigma^i_{SM}-\sigma^i_{NEW}}{\sigma^i_{SM}\Delta^i_{exp}}\right)^2
\end{eqnarray}
where
\begin{eqnarray}
\sigma^i_{SM}=\int_{z_i}^{z_{i+1}}\frac{d\sigma_{SM}}{dz}\,dz \\
\sigma^i_{NEW}=\int_{z_i}^{z_{i+1}}\frac{d\sigma_{NEW}}{dz}\,dz \\
\Delta^i_{exp}=\sqrt{{\delta^i_{stat.}}^2+{\delta^i_{syst.}}^2},\,\,\,\,\,\,
z=\cos\theta
\end{eqnarray}
Here, subscript "NEW" represents the cross section including
unparticle contributions. $\delta_{stat.}$ and $\delta_{syst.}$ are
the statistical and systematic errors. We have divided the range of
$\cos\theta$ into six equal pieces for the binning procedure and
have considered at least 100 events in each bin. We impose only a
pseudo-rapidity cut of $|\eta|<2.5$ which is necessary for the
central detector acceptance.

\begin{figure}
\includegraphics{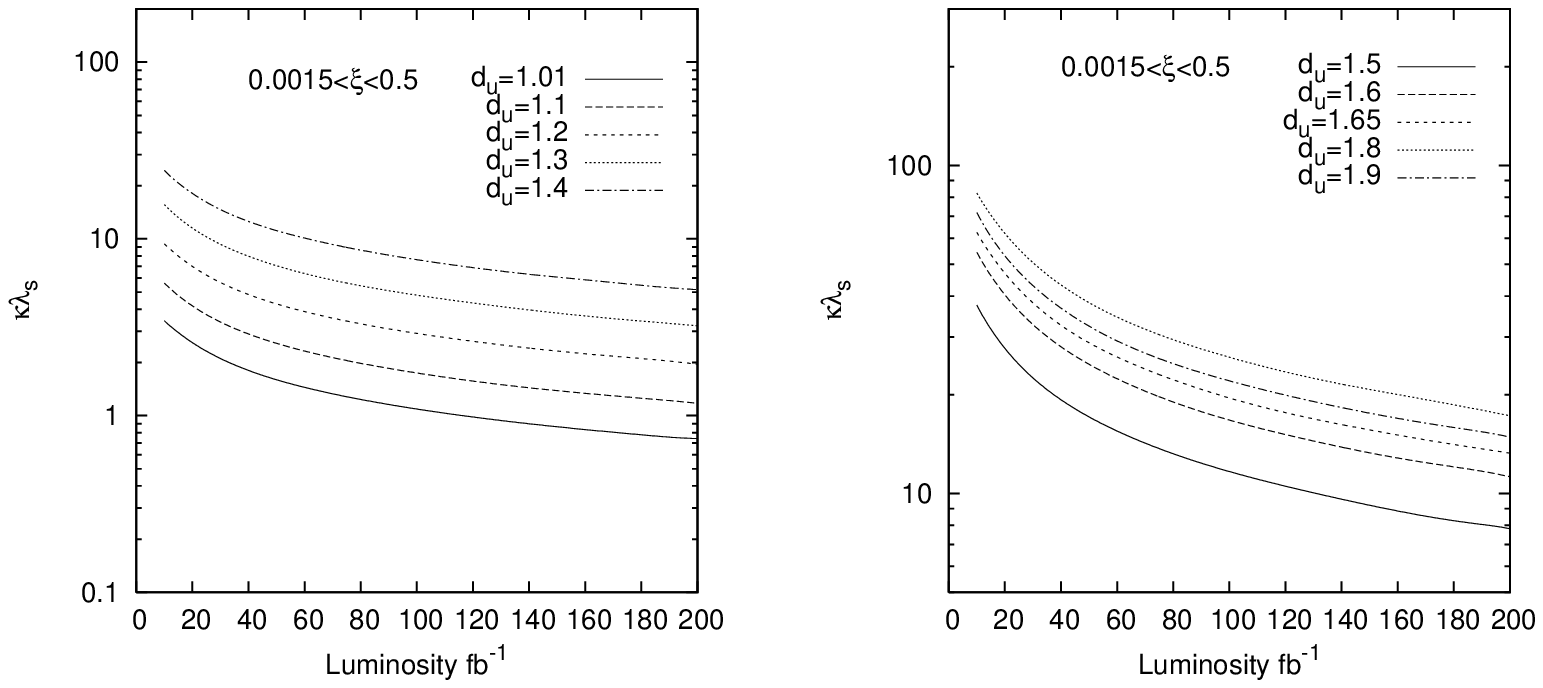}
\caption{95\% C.L sensitivity of $pp\to p\ell^-\ell^+p$ to the
product of scalar unparticle couplings $\kappa\lambda_S$ as a
function of integrated LHC luminosity for an acceptance of
$0.0015<\xi<0.5$. Various values of the scale dimension are stated
on the figures.
 We impose the cuts;
$|\vec{q}_{1t}+\vec{q}_{2t}|<30$MeV, $|\eta|<2.5$ and $p_t>460$GeV.
$\Lambda_U$ is taken to be 3 TeV. \label{fig4}}
\end{figure}

\begin{figure}
\includegraphics{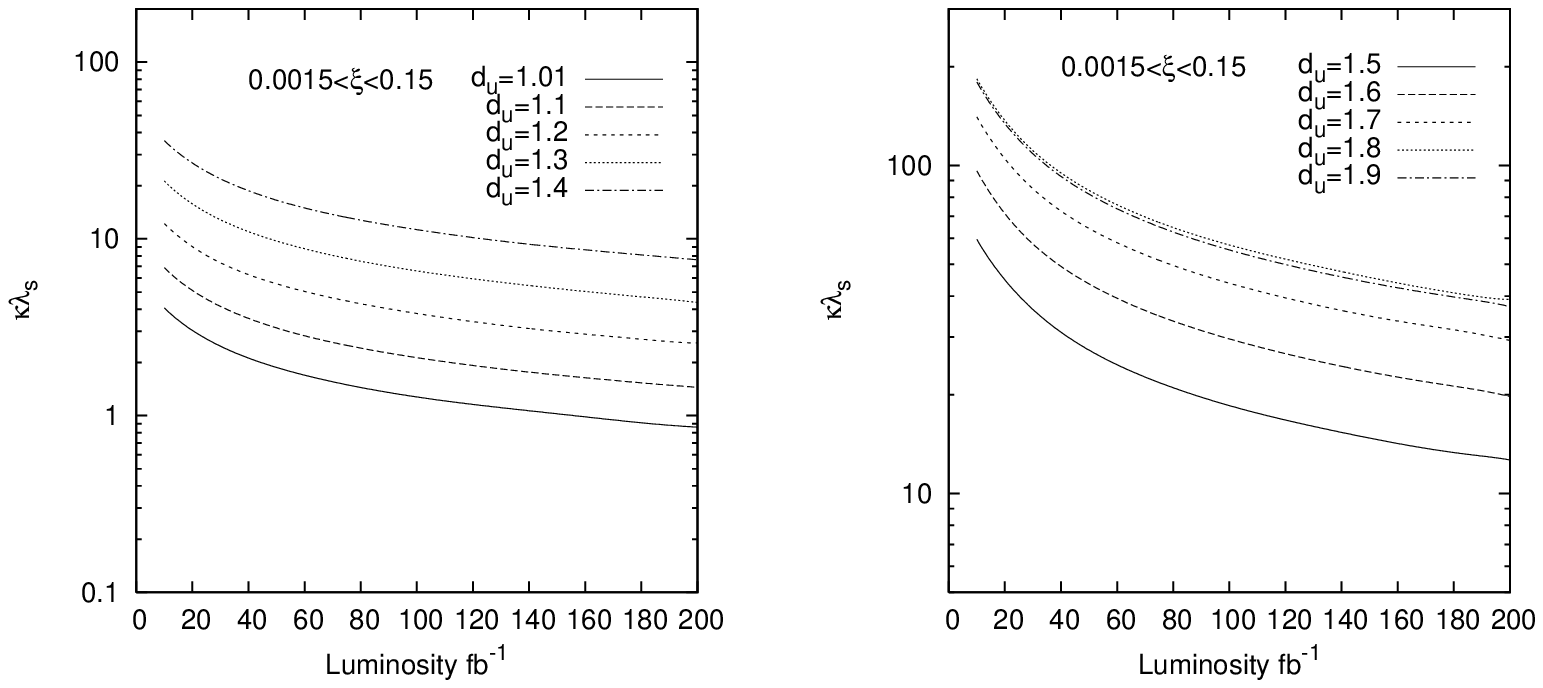}
\caption{95\% C.L sensitivity of $pp\to p\ell^-\ell^+p$ to the
product of scalar unparticle couplings $\kappa\lambda_S$ as a
function of integrated LHC luminosity for an acceptance of
$0.0015<\xi<0.15$. Various values of the scale dimension are stated
on the figures.
 We impose the cuts;
$|\vec{q}_{1t}+\vec{q}_{2t}|<30$MeV, $|\eta|<2.5$ and $p_t>420$GeV.
$\Lambda_U$ is taken to be 3 TeV. \label{fig5}}
\end{figure}

\begin{figure}
\includegraphics{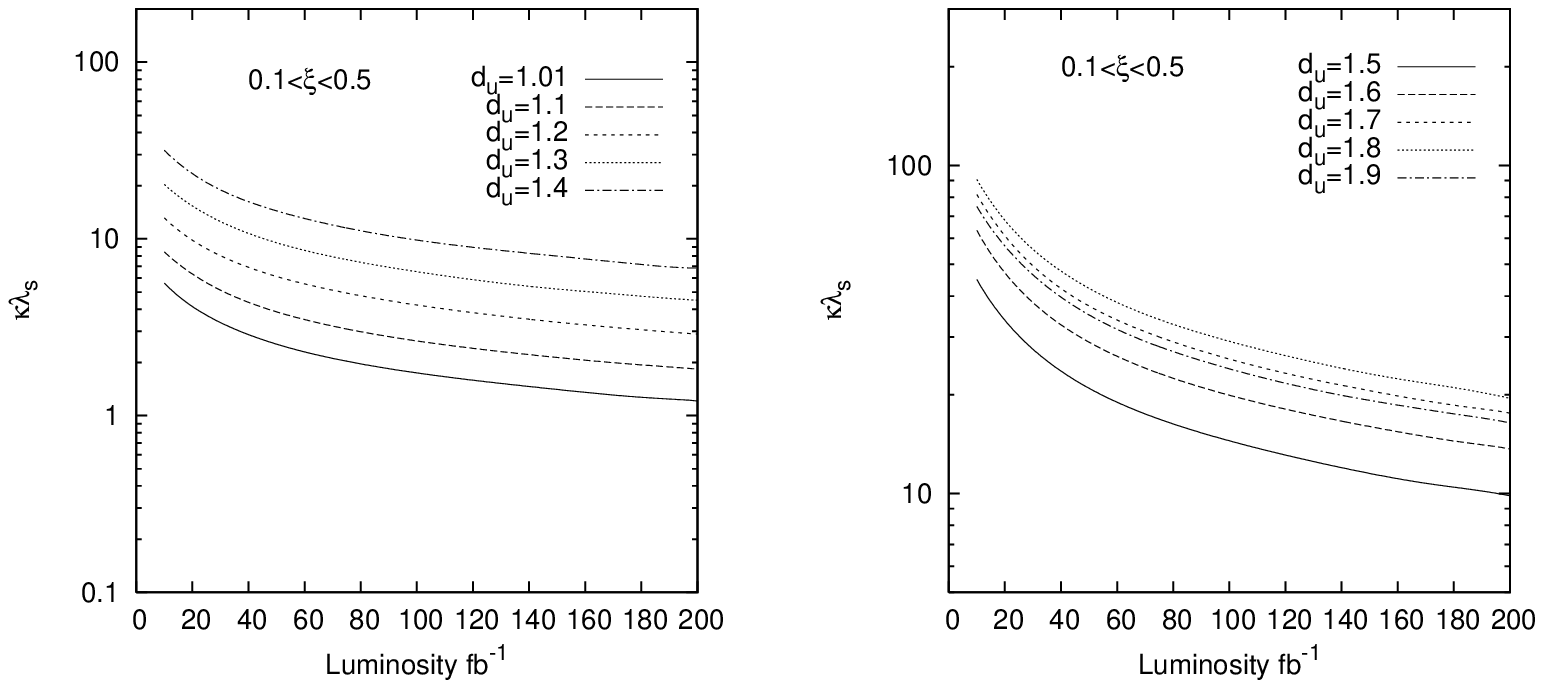}
\caption{95\% C.L sensitivity of $pp\to p\ell^-\ell^+p$ to the
product of scalar unparticle couplings $\kappa\lambda_S$ as a
function of integrated LHC luminosity for an acceptance of
$0.1<\xi<0.5$. Various values of the scale dimension are stated on
the figures.
 We impose the cuts;
$|\vec{q}_{1t}+\vec{q}_{2t}|<30$MeV, $|\eta|<2.5$ and $p_t>30$GeV.
$\Lambda_U$ is taken to be 3 TeV. \label{fig6}}
\end{figure}

\begin{figure}
\includegraphics{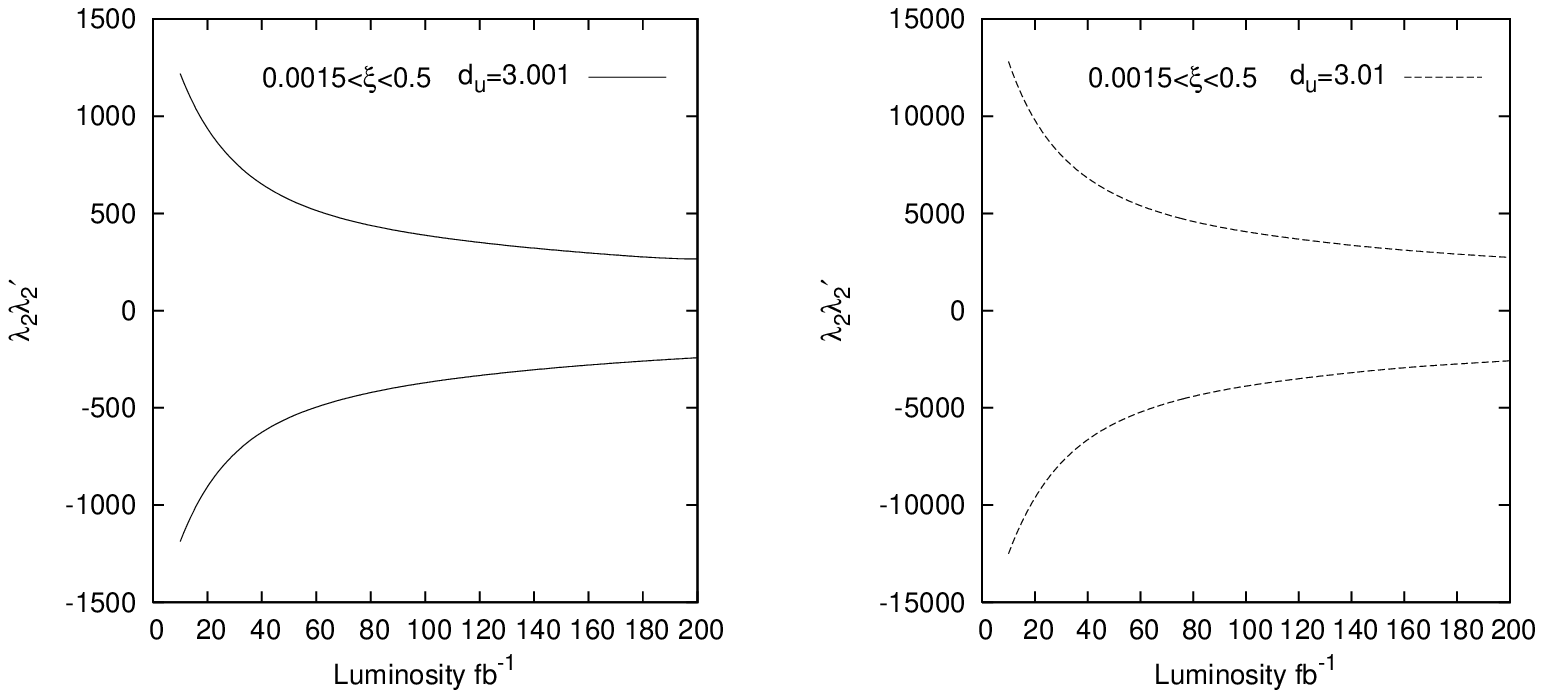}
\caption{The same as Fig.\ref{fig4} but for product of tensor
unparticle couplings $\lambda_2\lambda_2^\prime$. \label{fig7}}
\end{figure}

\begin{figure}
\includegraphics{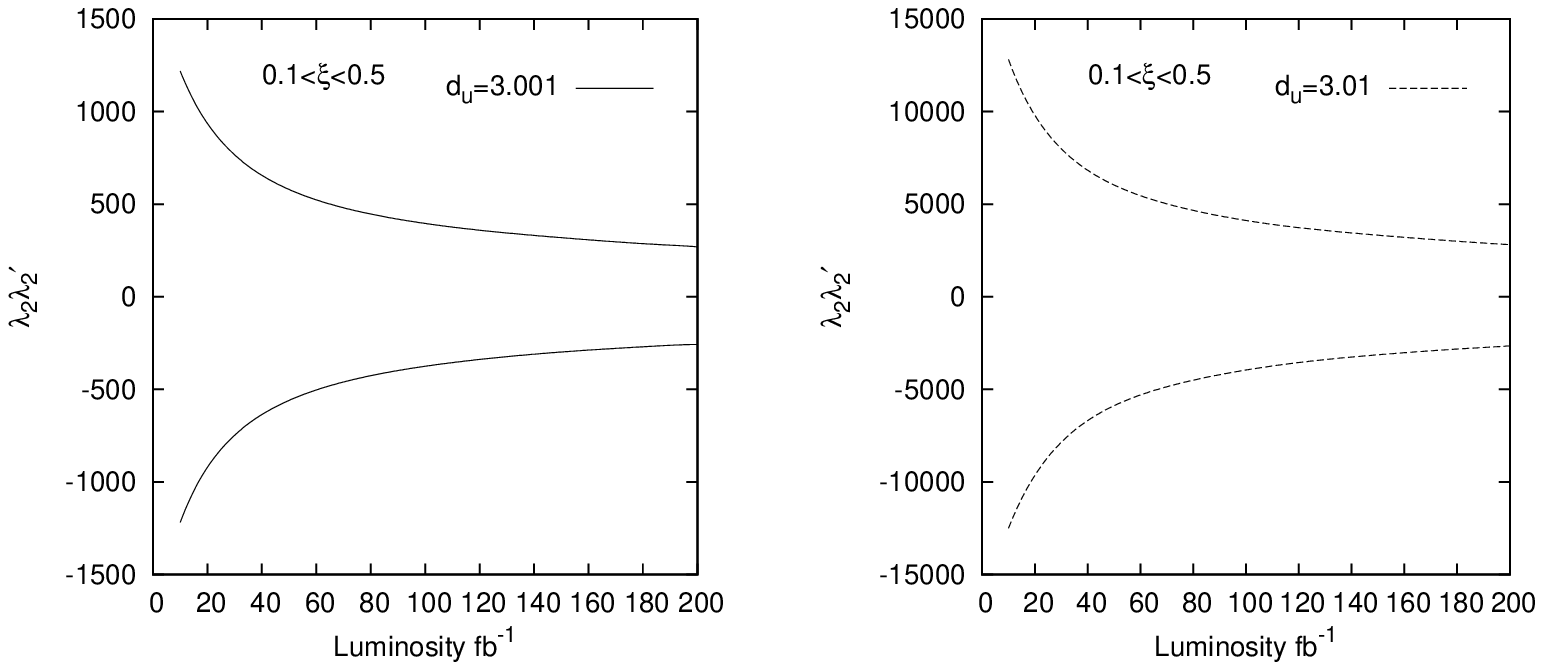}
\caption{The same as Fig.\ref{fig6} but for product of tensor
unparticle couplings $\lambda_2\lambda_2^\prime$.
 \label{fig8}}
\end{figure}

For a concrete result we have obtained 95\% confidence level (C.L.)
limits on the unparticle couplings. The number of observed events is
assumed to be equal to the SM prediction
$N_{obs}=0.9\,L\,\sigma_{SM}$ where $L$ is the integrated luminosity
and 0.9 is the QED two-photon survival probability
\cite{qedsurvival}. We assume that electrons and muons in the final
state can be observed in the central detectors with an acceptance
cut of  $|\eta|<2.5$. In Fig.\ref{fig4}-\ref{fig6} we present the
sensitivity of $pp\to p\ell^-\ell^+p$ to the product of scalar
unparticle couplings $\kappa\lambda_S$ from a Poisson distribution.
Sensitivity limits are given as a function of integrated LHC
luminosity for the acceptances of $0.0015<\xi<0.5$,
$0.0015<\xi<0.15$ and $0.1<\xi<0.5$. Since the scalar unparticle
contribution is symmetric in the negative and positive intervals of
the coupling we present our results only for positive
$\kappa\lambda_S$. We see from the figures that the decrease in
$d_U$ generally improves the sensitivity limits. The most sensitive
results are obtained at $d_U=1.01$. On the other hand, limits for
$d_U=1.9$ are sensitive than the limits for $d_U=1.8$. This is
reasonable from $\sin^2(d_U\pi)$ dependence of the denominator of
the scalar unparticle contribution (\ref{ggll}). Limits on the
tensor unparticle couplings are given in Fig.\ref{fig7} and
Fig.\ref{fig8} from a Poisson distribution for the acceptances
$0.0015<\xi<0.5$ and $0.1<\xi<0.5$ respectively. Limits for
$0.0015<\xi<0.15$ are too weak compared with other cases so we do
not plot them. We see from the figures that limits for
$0.0015<\xi<0.5$ and $0.1<\xi<0.5$ cases are almost the same. This
originates from the fact that at low energies tensor unparticle
contribution is very suppressed and the main contribution comes from
high energy region.

In Fig.\ref{fig4}-\ref{fig6}, $p_t$ cuts on the final leptons are
proposed considering a luminosity of 200 $fb^{-1}$. On the other
hand, these cuts are not the optimum ones for other luminosity
values. For a given luminosity, limits on the unparticle couplings
can be improved by adjusting the $p_t$ cut on the final leptons. To
this purpose, we present Table \ref{tab1} and Table \ref{tab2} where
we take into account different $p_t$ cuts for different
luminosities. We show that especially for small luminosity values,
considerable improvement is obtained in the limits by adjusting the
$p_t$ cut. In Table \ref{tab3} and Table \ref{tab4} we show 95\%C.L.
lower bounds on the energy scale $\Lambda_U$ with the same
luminosity values and $p_t$ cuts of Tables \ref{tab1}-\ref{tab2}. In
the tables the couplings are taken to be $\kappa$=$\lambda_S$=1 and
$\lambda_{PS}=\lambda_2=\lambda_2^\prime=0$.

\begin{figure}
\includegraphics{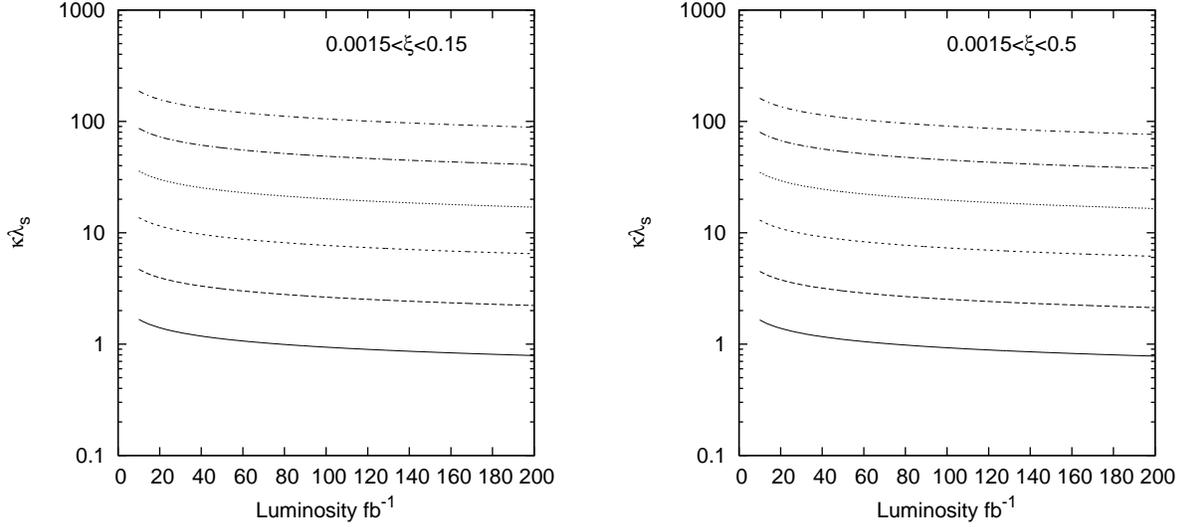}
\caption{95\% C.L sensitivity of $pp\to p\ell^-\ell^+p$ to the
product of scalar unparticle couplings $\kappa\lambda_S$ as a
function of integrated LHC luminosity for the acceptances
$0.0015<\xi<0.15$ (left panel) and $0.0015<\xi<0.5$ (right panel).
Limits are estimated using a simple $\chi^2$ test from angular
distribution without a systematic error. Curves from bottom to top
correspond to increasing values of
$d_U=1.01,\,1.1,\,1.2,\,1.3,\,1.4$ and 1.5. We impose the cuts;
$|\vec{q}_{1t}+\vec{q}_{2t}|<30$MeV, $|\eta|<2.5$ and $\Lambda_U$ is
taken to be 3 TeV. \label{fig9}}
\end{figure}

\begin{figure}
\includegraphics{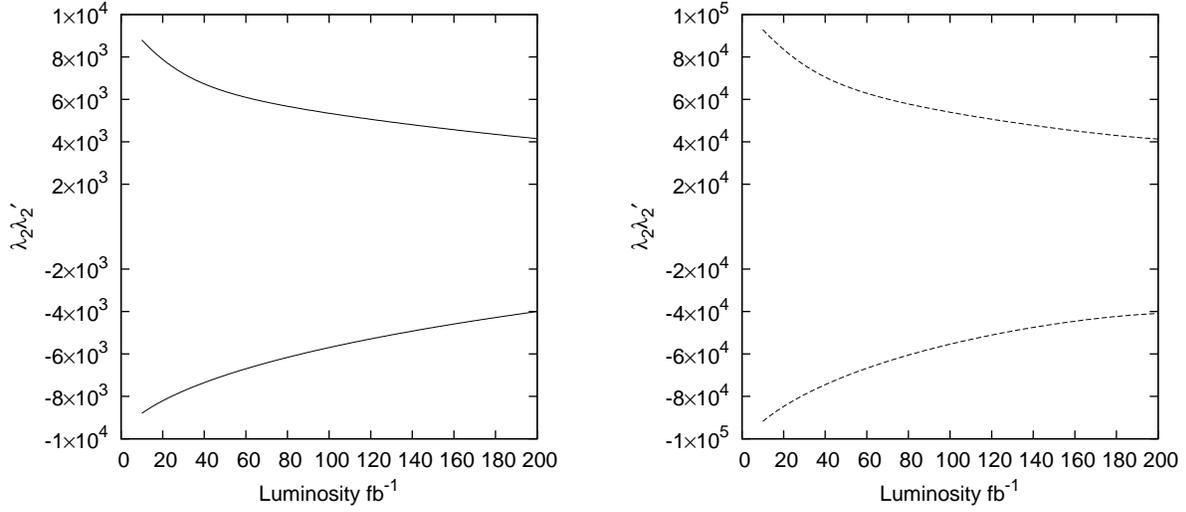}
\caption{95\% C.L sensitivity of $pp\to p\ell^-\ell^+p$ to the
product of tensor unparticle couplings $\lambda_2\lambda_2^\prime$
as a function of integrated LHC luminosity for the acceptance
$0.0015<\xi<0.5$. Limits are estimated using a simple $\chi^2$ test
from angular distribution without a systematic error. Solid line
corresponds to the limit for $d_U=3.001$ (left panel) and the dotted
line corresponds to $d_U=3.01$ (right panel). We impose the cuts;
$|\vec{q}_{1t}+\vec{q}_{2t}|<30$MeV, $|\eta|<2.5$ and $\Lambda_U$ is
taken to be 3 TeV. \label{fig10}}
\end{figure}

In Fig.\ref{fig9} and  Fig.\ref{fig10}, we estimate 95\% C.L. limits
for scalar and tensor unparticle couplings using a simple $\chi^2$
test without a systematic error. We do not estimate the limits for
$0.1<\xi<0.5$ case since the SM cross section is about $1.1\times
10^{-6}$ pb. Therefore number of SM event is smaller than 1 even for
a luminosity of $200fb^{-1}$. We see from Fig.\ref{fig9} and
Fig.\ref{fig10} that limits rapidly get worse as the $d_U$
increases. This behavior is common in the analysis from a Poisson
distribution but deterioration rate is high in the $\chi^2$ case.
For example, when the detector acceptance is $0.0015<\xi<0.5$,
limits on the scalar unparticle couplings from Poisson distribution
deteriorated by a factor of 7 as the $d_U$ increases from 1.01 to
1.4. But this factor is approximately 40 in the $\chi^2$ analysis.
Therefore, $\chi^2$ analysis from angular distribution is favorable
for small values of the scale dimension close to unity. Hence we do
not give the limits on the scalar unparticle couplings for $d_U
>1.5$. But for comparison, we also give the limits on the tensor
unparticle couplings for $d_U=3.001$ and 3.01 (Fig.\ref{fig10}).

\subsection{Exclusive two photon production}

\begin{figure}
\includegraphics{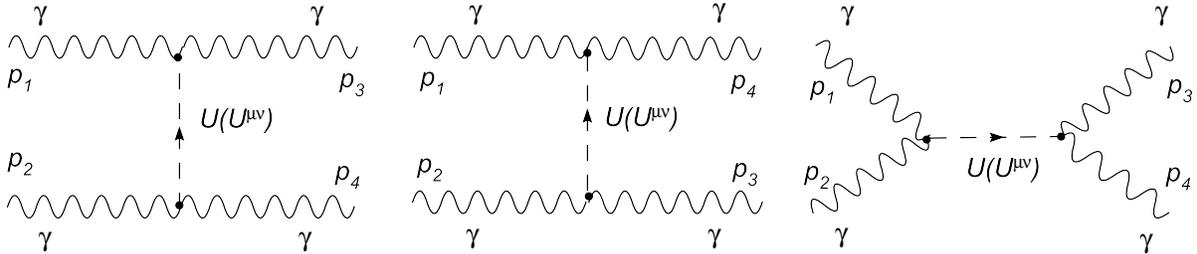}
\caption{Tree-level Feynman diagrams for the subprocess
$\gamma\gamma \to \gamma\gamma$. \label{fig11}}
\end{figure}

The subprocess $\gamma\gamma \to \gamma\gamma$ is absent in the SM
at the tree-level. Scalar and tensor unparticles contribute to the
process through t, u and s-channel diagrams (Fig.\ref{fig11}). The
polarization summed scattering amplitude for Fig.\ref{fig11} is
given by \cite{chang}

\begin{eqnarray}
\label{gggg}|M|^2=&&\frac{
A_{d_U}^2}{\sin^2{(d_U\pi)}}\left\{\frac{16\kappa^4}{\Lambda_U^{4d_U}}\left[
|t|^{2d_U}+s^{2d_U}+|u|^{2d_U}+|t|^{d_U}|u|^{d_U}\right.\right.\nonumber
\\ \nonumber
&&\left.\left.+\cos(d_U\pi)\left(|t|^{d_U}s^{d_U}+s^{d_U}|u|^{d_U}\right)\right]
+\frac{{\lambda_2^\prime}^4}{2\Lambda_U^{4d_U}}\left[s^{2d_U-4}(t^4+u^4)
\right.\right.\\ \nonumber
&&\left.\left.+{|t|}^{2d_U-4}(s^4+u^4)+{|u|}^{2d_U-4}(s^4+t^4)+2{|t|}^{d_U-2}
{|u|}^{d_U-2}s^4\right.\right. \\  &&\left.\left.
+2\cos(d_U\pi)s^{d_U-2}({|t|^{d_U-2}}u^4+{|u|}^{d_U-2}t^4)\right]\right\}
\end{eqnarray}

When we compare this amplitude with the amplitude of $\gamma\gamma
\to \ell^-\ell^+$, we see that unlike from amplitude (\ref{ggll}), t
and u-channel scalar unparticle exchange interfere with the
s-channel one. The prominent advantage of the subprocess
$\gamma\gamma \to \gamma\gamma$ is that it isolates the couplings
$\kappa$ and $\lambda_2^\prime$. As we have seen, this is not the
case in $\gamma\gamma \to \ell^-\ell^+$.

\begin{figure}
\includegraphics{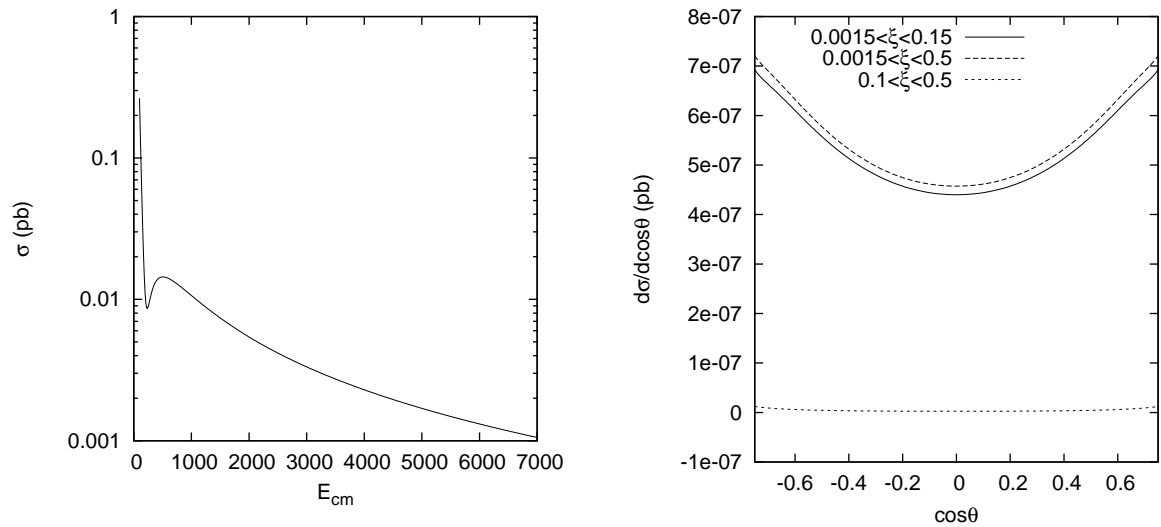}
\caption{Figure on the left shows total SM 1-loop contribution as a
function of center of mass energy of the two photon system. Figure
on the right shows angular distribution of the total SM 1-loop
contributions for various forward detector acceptances stated on the
figure. We impose a cut of $|\cos \theta|<0.86$ in the left panel.
\label{fig12}}
\end{figure}

SM background is originated from loop diagrams involving
contributions from charged fermions and W bosons. In Fig.\ref{fig12}
we plot SM expectation at 1-loop level using the form factors from
Refs.\cite{jikia,gounaris1,gounaris2}. We observe from the left
panel of Fig.\ref{fig12} that the SM cross section rapidly grows
when the energy decreases from approximately 200 GeV. This behavior
originates from fermion loop contributions. We impose a cut of
$\sqrt {s_{\gamma\gamma}}> 250$ GeV on the invariant mass of final
photons to reduce the contribution coming from fermion loops. (This
constraint is automatically satisfied in $0.1<\xi<0.5$ since
$E_{min}=1400$ GeV.) Furthermore we impose a pseudo-rapidity cut of
$|\eta|<0.88$ for the cases; $0.0015<\xi<0.5$ and $0.0015<\xi<0.15$.
These cuts effectively suppress SM loop contributions coming from
fermion and W loops but do not spoil unparticle limits more than a
factor of 1.5. This is reasonable since the main unparticle
contribution comes from high energy region and it does not peak in
the forward and backward directions. On the other hand, we see from
the right panel of Fig.\ref{fig12} that SM cross section peaks in
the forward and backward directions. Total SM cross sections from W
and fermion loops are; $3.50\times10^{-7}$pb for $0.0015<\xi<0.15$
and $3.52\times10^{-7}$pb for $0.0015<\xi<0.5$ when the mentioned
cuts in place. Therefore they are negligible even with a luminosity
of 200$fb^{-1}$. In $0.1<\xi<0.5$ case with $|\eta|<2.5$ loop
contributions are much more smaller. Total SM cross section is
$2.60\times10^{-8}$pb.

\begin{figure}
\includegraphics{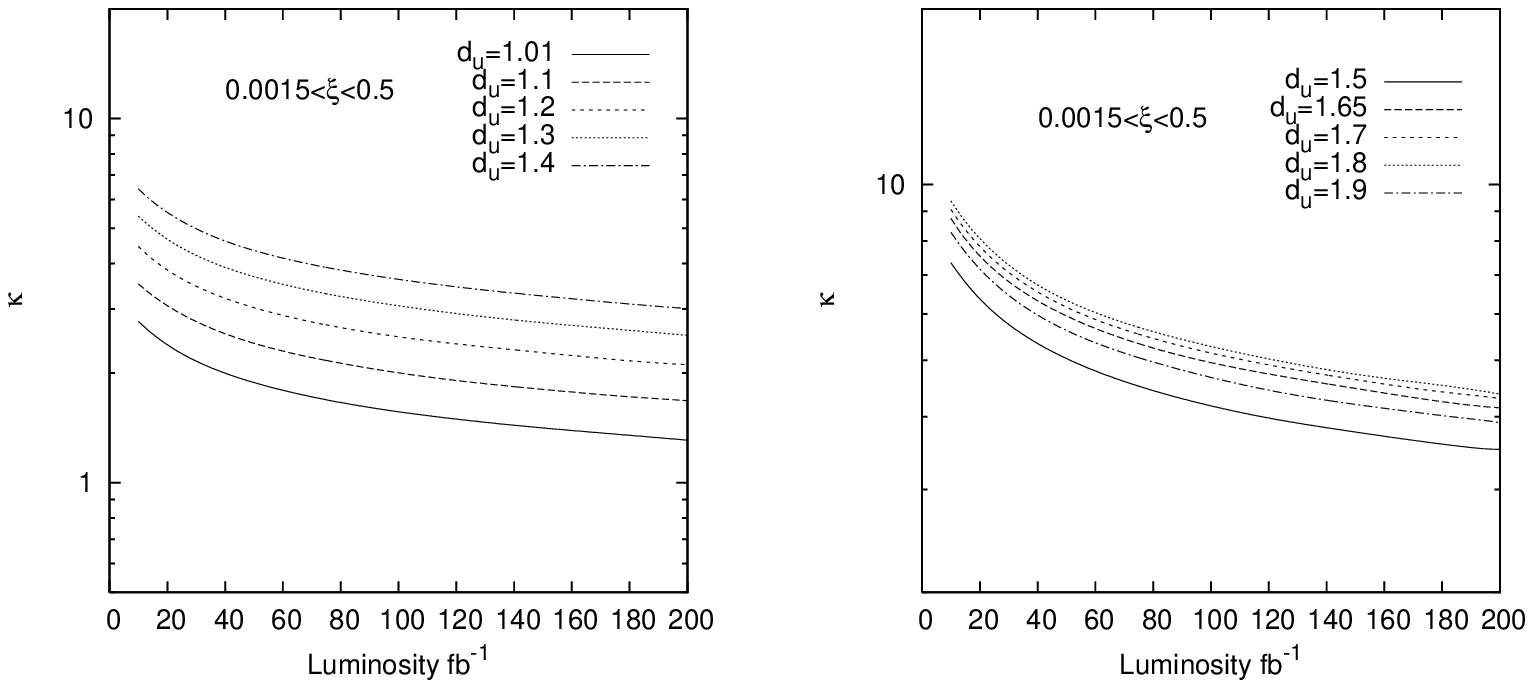}
\caption{95\% C.L sensitivity of $pp\to p\gamma\gamma p$ to scalar
unparticle coupling $\kappa$ as a function of integrated LHC
luminosity for an acceptance of $0.0015<\xi<0.5$. Various values of
the scale dimension are stated on the figures. We impose the cuts;
$|\vec{q}_{1t}+\vec{q}_{2t}|<30$MeV, $\sqrt{s_{\gamma\gamma}}>250$
GeV and $|\eta|<0.88$. $\Lambda_U$ is taken to be 3
TeV.\label{fig13}}
\end{figure}

\begin{figure}
\includegraphics{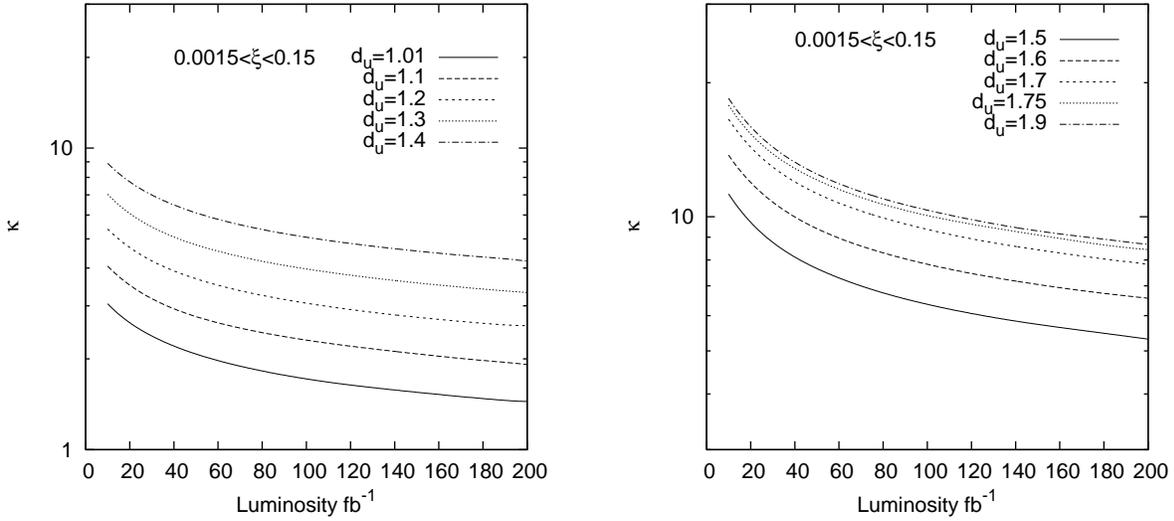}
\caption{The same as Fig.\ref{fig13} but for forward detector
acceptance $0.0015<\xi<0.15$.\label{fig14}}
\end{figure}

\begin{figure}
\includegraphics{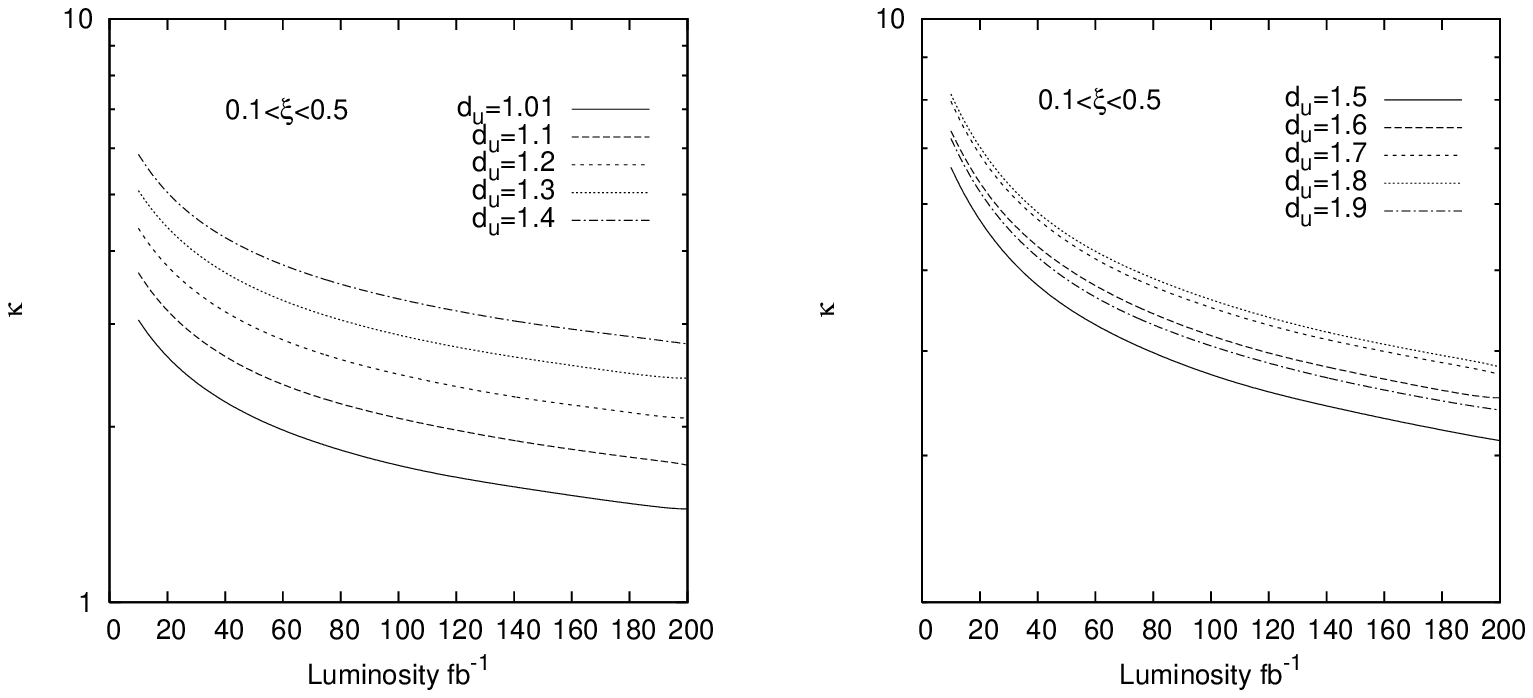}
\caption{95\% C.L sensitivity of $pp\to p\gamma\gamma p$ to scalar
unparticle coupling $\kappa$ as a function of integrated LHC
luminosity for an acceptance of $0.1<\xi<0.5$. Various values of the
scale dimension are stated on the figures. We impose the cuts;
$|\vec{q}_{1t}+\vec{q}_{2t}|<30$MeV and $|\eta|<2.5$. $\Lambda_U$ is
taken to be 3 TeV.\label{fig15}}
\end{figure}

\begin{figure}
\includegraphics{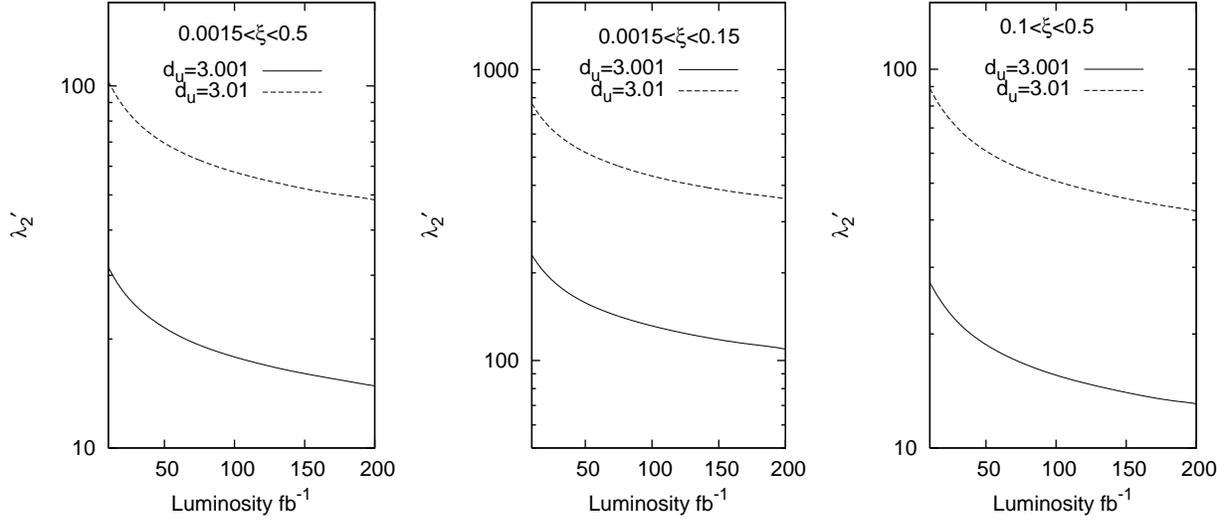}
\caption{95\% C.L sensitivity of $pp\to p\gamma\gamma p$ to tensor
unparticle coupling $\lambda_2^\prime$ as a function of integrated
LHC luminosity. Different panels show different detector
acceptances. The solid lines are for scale dimension $d_U=3.001$ and
dotted lines are for $d_U=3.01$. We impose the cuts;
$|\vec{q}_{1t}+\vec{q}_{2t}|<30$MeV, $\sqrt{s_{\gamma\gamma}}>250$
GeV and $|\eta|<0.88$ for the acceptances $0.0015<\xi<0.5$ and
$0.0015<\xi<0.15$ and we impose $|\vec{q}_{1t}+\vec{q}_{2t}|<30$MeV
and $|\eta|<2.5$ for $0.1<\xi<0.5$. $\Lambda_U$ is taken to be 3
TeV. \label{fig16}}
\end{figure}

In order to obtain more realistic results, we take into account a
photon efficiency of 90\% for each final photons in the numerical
calculations \cite{colas,baker}. In Fig.\ref{fig13}-\ref{fig15} we
plot sensitivity of $pp\to p\gamma\gamma p$ to scalar unparticle
coupling $\kappa$ as a function of integrated LHC luminosity for the
acceptances of $0.0015<\xi<0.5$, $0.0015<\xi<0.15$ and
$0.1<\xi<0.5$. The most sensitive results are obtained in
$0.0015<\xi<0.5$ for $d_U=1.01$ and 1.1. But $0.1<\xi<0.5$ case
gives better limits for $d_U>1.1$. In Fig.\ref{fig16} we present the
limits on tensor unparticle coupling $\lambda_2^\prime$ for the
scale dimensions $d_U=3.001$ and $d_U=3.01$. We observe from the
figure that the limits for $0.0015<\xi<0.5$ and $0.1<\xi<0.5$ cases
are close to each other. Therefore forward detectors with
acceptances $0.0015<\xi<0.5$ and $0.1<\xi<0.5$ have almost same
potential to probe tensor unparticle contribution through the
process $pp\to p\gamma\gamma p$.

\begin{figure}
\includegraphics{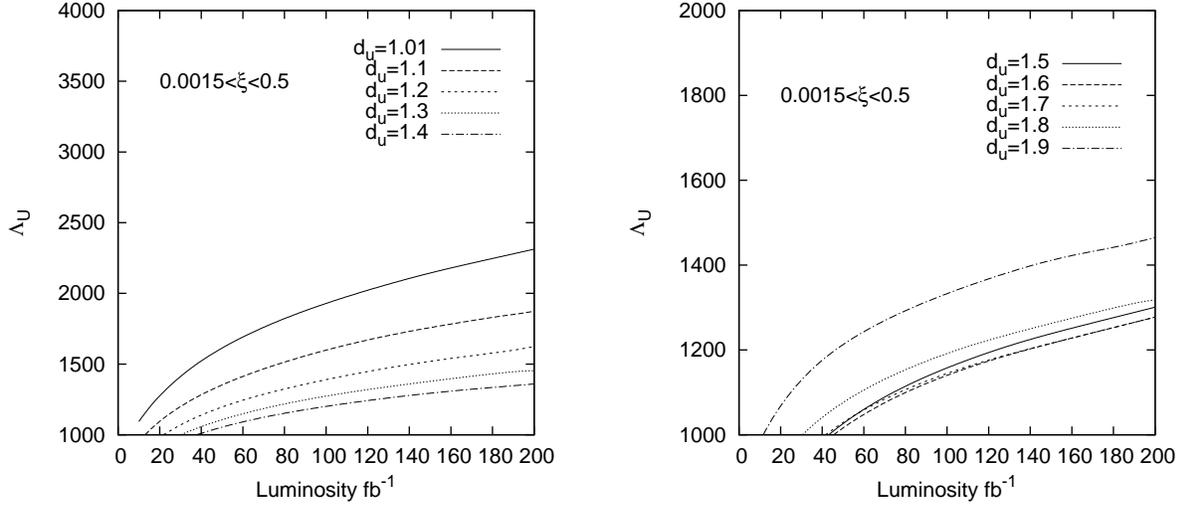}
\caption{95\% C.L. lower bounds on the energy scale $\Lambda_U$ as a
function of integrated LHC luminosity for $pp\to p\gamma\gamma p$.
Various values of the scale dimension are stated on the figures. The
couplings are taken to be $\kappa$=1 and $\lambda_2^\prime=0$.
Limits of $\Lambda_U$ are given in units of GeV. Forward detector
acceptance is $0.0015<\xi<0.5$ and we impose the cuts;
$|\vec{q}_{1t}+\vec{q}_{2t}|<30$MeV, $\sqrt{s_{\gamma\gamma}}>250$
GeV and $|\eta|<0.88$. \label{fig17}}
\end{figure}

Lower bounds on the energy scale $\Lambda_U$ are obtained as a
function of integrated LHC luminosity in Fig.\ref{fig17}. A
comparison with Table.\ref{tab3} shows that $pp\to p\ell^-\ell^+p$
is more sensitive to $\Lambda_U$ for the values of the scale
dimension which are close to unity such as $d_U=1.01$ and $d_U=1.1$.
On the other hand, the process $pp\to p\gamma\gamma p$ is more
sensitive to the energy scale for the values of the scale dimension
between $1.4 - 1.9$.

\section{Conclusions}

Current experimental restrictions on unparticle couplings were
widely studied in the literature. Although there still remains some
reactions which was not examined, task to find current experimental
limits is almost completed. LHC has started operating and its
potential to probe unparticles has been under research
\cite{mathews,mohanta,rizzo1,rizzo2,rizzo3,alan,kumar1,feng1,ghosh,kumar2,kumar3,arai}.
Limits from LHC have been provided by two photon production via
$gg,q\bar{q} \to \gamma\gamma$ \cite{feng1}. A comparison of our
limits with the limits of $gg,q\bar{q} \to \gamma\gamma$ is not
possible in general since the reactions involve different type of
couplings. If we assume that unparticle couplings to quarks and
gluons are equal to its couplings to leptons and photons then we
conclude that our limits are weaker than the limits obtained through
these reactions at the LHC but stronger than the limits obtained at
the Tevatron \cite{feng1}. On the other hand, exclusive
$\ell^-\ell^+$ and $\gamma\gamma$ production through $\gamma\gamma$
fusion provide very clean environment due to absence of the proton
remnants. Therefore any signal which conflicts with the SM
predictions would be a convincing evidence for new physics.

The exclusive two photon production $pp\to p\gamma\gamma p$ isolates
the couplings $\gamma\gamma U$, $\gamma\gamma U^{\mu\nu}$. This is a
prominent advantage of $pp\to p\gamma\gamma p$ and it can not be
achieved in any other process at the LHC. In the future,
$\gamma\gamma$ colliders are expected to be designed complementary
to linear $e^+e^-$ colliders \cite{tesla}. At the $\gamma\gamma$
mode of a linear collider, $\gamma\gamma U$ and $\gamma\gamma
U^{\mu\nu}$ couplings can be probed with a high precision
\cite{chang,okada,cakir}.

The process $pp\to p\ell^-\ell^+p$ was proposed as a luminosity
monitor for the LHC \cite{lhc3}. If it is used to measure luminosity
then it is important to know its sensitivity to new physics for a
given acceptance range. We have explored sensitivity of $pp\to
p\ell^-\ell^+p$ to unparticles with three different forward detector
acceptances. We show that $0.1<\xi<0.5$ case is least sensitive to
scalar unparticles for $d_U=1.01 - 1.2$ but $0.0015<\xi<0.15$ case
is least sensitive for $d_U=1.3 - 1.9$. Tensor unparticle
contribution rapidly grows with energy. Forward detector acceptance
of $0.0015<\xi<0.15$ is least sensitive to tensor unparticle
contribution.

\pagebreak

\pagebreak


\pagebreak

\begin{table}
\caption{Sensitivity of $pp\to p\ell^-\ell^+p$ to the product of
scalar unparticle couplings $\kappa\lambda_S$ at 95\% C.L. for
various values of the scale dimension $d_U$ and integrated LHC
luminosities. We impose different cuts on the transverse momentum of
final leptons for different luminosity values. Forward detector
acceptance is $0.0015<\xi<0.5$ and $\Lambda_U$ is taken to be 3 TeV.
\label{tab1}}
\begin{ruledtabular}
\begin{tabular}{ccccc}
Luminosity& 10$fb^{-1}$ &50$fb^{-1}$ &100$fb^{-1}$&200$fb^{-1}$ \\
 $p_{t,min}$&210 GeV &320 GeV &380 GeV &460 GeV \\
\hline
$d_U=1.01$ &(-1.4,\,1.4) &(-1.0,\,1.0) &(-0.8,\,0.8) &(-0.7,\,0.7) \\
$d_U=1.1$  &(-2.5,\,2.5) &(-1.6,\,1.6) &(-1.4,\,1.4) &(-1.2,\,1.2)   \\
$d_U=1.2$  &(-4.8,\,4.8) &(-2.9,\,2.9) &(-2.3,\,2.3) &(-2.0,\,2.0) \\
$d_U=1.3$  &(-8.8,\,8.8) &(-5.0,\,5.0) &(-3.9,\,3.9) &(-3.2,\,3.2) \\
$d_U=1.4$  &(-15.6,\,15.6) &(-8.1,\,8.1) &(-6.4,\,6.4) &(-5.2,\,5.2) \\
$d_U=1.5$  &(-25.6,\,25.6) &(-13.1,\,13.1) &(-10.0,\,10.0) &(-7.8,\,7.8) \\
$d_U=1.6$  &(-40.6,\,40.6) &(-19.7,\,19.7) &(-15.0,\,15.0) &(-11.3,\,11.3) \\
$d_U=1.7$  &(-56.9,\,56.9) &(-27.5,\,27.5) &(-19.7,\,19.7) &(-15.2,\,15.2) \\
$d_U=1.8$  &(-67.8,\,67.8) &(-32.5,\,32.5) &(-23.4,\,23.4) &(-17.3,\,17.3) \\
$d_U=1.9$  &(-61.3,\,61.3) &(-27.5,\,27.5) &(-20.2,\,20.2) &(-14.8,\,14.8) \\
\end{tabular}
\end{ruledtabular}
\end{table}

\begin{table}
\caption{The same as table \ref{tab1} but for $0.0015<\xi<0.15$.
\label{tab2}}
\begin{ruledtabular}
\begin{tabular}{ccccc}
Luminosity& 10$fb^{-1}$ &50$fb^{-1}$ &100$fb^{-1}$&200$fb^{-1}$ \\
 $p_{t,min}$&200 GeV &310 GeV &360 GeV &420 GeV \\
\hline
$d_U=1.01$ &(-1.5,\,1.5) &(-1.1,\,1.1) &(-1.0,\,1.0) &(-0.9,\,0.9) \\
$d_U=1.1$  &(-2.7,\,2.7) &(-2.0,\,2.0) &(-1.6,\,1.6) &(-1.5,\,1.5) \\
$d_U=1.2$  &(-5.3,\,5.3) &(-3.6,\,3.6) &(-3.0,\,3.0) &(-2.6,\,2.6) \\
$d_U=1.3$  &(-10.0,\,10.0) &(-6.4,\,6.4) &(-5.3,\,5.3) &(-4.4,\,4.4) \\
$d_U=1.4$  &(-18.8,\,18.8) &(-11.3,\,11.3) &(-9.1,\,9.1) &(-7.6,\,7.6) \\
$d_U=1.5$  &(-33.8,\,33.8) &(-19.4,\,19.4) &(-15.6,\,15.6) &(-12.7,\,12.7) \\
$d_U=1.6$  &(-57.5,\,57.5) &(-31.3,\,31.3) &(-25.0,\,25.0) &(-19.8,\,19.8) \\
$d_U=1.7$  &(-90.6,\,90.6) &(-47.5,\,47.5) &(-37.5,\,37.5) &(-29.3,\,29.3) \\
$d_U=1.8$  &(-122.5,\,122.5) &(-65.0,\,65.0) &(-48.8,\,48.8) &(-39.1,\,39.1) \\
$d_U=1.9$  &(-122.5,\,122.5) &(-62.5,\,62.5) &(-48.7,\,48.7) &(-37.1,\,37.1) \\
\end{tabular}
\end{ruledtabular}
\end{table}

\begin{table}
\caption{Sensitivity of $pp\to p\ell^-\ell^+p$ to $\Lambda_U$ at
95\% C.L. for various values of the scale dimension $d_U$ and
integrated LHC luminosities. We impose different cuts on the
transverse momentum of final leptons for different luminosity
values. The couplings are taken to be $\kappa$=$\lambda_S$=1 and
$\lambda_{PS}=\lambda_2=\lambda_2^\prime=0$. Lower bounds of
$\Lambda_U$ are given in units of GeV. Forward detector acceptance
is $0.0015<\xi<0.5$. \label{tab3}}
\begin{ruledtabular}
\begin{tabular}{ccccc}
Luminosity& 10$fb^{-1}$ &50$fb^{-1}$ &100$fb^{-1}$&200$fb^{-1}$ \\
 $p_{t,min}$&210 GeV &320 GeV &380 GeV &460 GeV \\
\hline
$d_U=1.01$ &2109 &3125 &3625 &4063 \\
$d_U=1.1$  &1367 &2000 &2375 &2625   \\
$d_U=1.2$  &977 &1406 &1625 &1875 \\
$d_U=1.3$  &781 &1094 &1281 &1438 \\
$d_U=1.4$  &654 &922 &1063 &1219 \\
$d_U=1.5$  &596 &828 &953 &1063 \\
$d_U=1.6$  &557 &773 &875 &1000 \\
$d_U=1.7$  &557 &758 &859 &969 \\
$d_U=1.8$  &586 &789 &891 &1000 \\
$d_U=1.9$  &693 &914 &1031 &1141 \\
\end{tabular}
\end{ruledtabular}
\end{table}

\begin{table}
\caption{The same as table \ref{tab3} but for $0.0015<\xi<0.15$.
\label{tab4}}
\begin{ruledtabular}
\begin{tabular}{ccccc}
Luminosity& 10$fb^{-1}$ &50$fb^{-1}$ &100$fb^{-1}$&200$fb^{-1}$ \\
 $p_{t,min}$&200 GeV &310 GeV &360 GeV &420 GeV \\
\hline
$d_U=1.01$ &2063 &2734 &3125 &3469 \\
$d_U=1.1$  &1313 &1734 &1992 &2203  \\
$d_U=1.2$  &906 &1219 &1367 &1547 \\
$d_U=1.3$  &703 &938 &1063 &1172 \\
$d_U=1.4$  &594 &773 &875 &969 \\
$d_U=1.5$  &516 &680 &766 &844 \\
$d_U=1.6$  &477 &625 &703 &766 \\
$d_U=1.7$  &461 &594 &664 &734 \\
$d_U=1.8$  &469 &609 &670 &739 \\
$d_U=1.9$  &533 &680 &750 &825 \\
\end{tabular}
\end{ruledtabular}
\end{table}

\end{document}